\documentclass[aps,prl,twocolumn,superscriptaddress,showpacs]{revtex4}
\usepackage{graphics}
\usepackage{bm}

\usepackage{amsmath}    
\usepackage{amssymb}    
\usepackage{graphicx}   
\usepackage{epsf}
\usepackage{epsfig}
\begin{document}

\title {Planar Josephson Tunnel Junctions in a Transverse Magnetic Field}
\thanks{Journal of Applied Physics vol.102, 093911 (2007)}

\author{R.\ Monaco}
\affiliation{Istituto di Cibernetica del C.N.R., 80078, Pozzuoli, Italy and Unita' INFM-Dipartimento di Fisica, Universita' di Salerno, 84081 Baronissi, Italy.}\email
{roberto@sa.infn.it}
\author{M.\ Aaroe}
\affiliation{Department of Physics, B309, Technical University of
Denmark, DK-2800 Lyngby, Denmark.} \email{aaroe@fysik.dtu.dk}
\author{V.\ P.\ Koshelets}
\affiliation{Institute of Radio Engineering and Electronics, Russian
Academy of Science, Mokhovaya 11, Bldg 7, 125009, Moscow,
Russia.}\email{valery@hitech.cplire.ru}
\author{J.\ Mygind}
\affiliation{Department of Physics, B309,  Technical University of
Denmark, DK-2800 Lyngby, Denmark.} \email{myg@fysik.dtu.dk}

\date{\today}

\begin{abstract}
Traditionally, since the discovery of the Josephson effect in 1962,
the magnetic diffraction pattern of planar Josephson tunnel
junctions has been recorded with the field applied in the plane of
the junction. Here we discuss the static junction properties in a
\emph{transverse} magnetic field where demagnetization effects
imposed by the tunnel barrier and electrodes geometry are important.
Measurements of the junction critical current versus magnetic field
in planar Nb-based high-quality junctions with different geometry,
size and critical current density show that it is advantageous to
use a transverse magnetic field rather than an in-plane field. The
conditions under which this occurs are discussed.
\end{abstract}


\maketitle

\section{Introduction}

It is well known that a magnetic field ${\bf H}$  modulates the
critical current $I_c$ of a Josephson Tunnel Junction (JTJ)
\cite{barone}. Indeed the occurrence of such diffraction phenomena
$I_c (H)$ is one of the most striking behaviors of
JTJs\cite{anderson}. In the case of a planar JTJ, it was Josephson
himself \cite{brian} who pointed out that the gradient of the
Josephson phase $\phi$, which is the difference between the complex
wavefunction phases in the electrodes, can be expressed as

\begin{equation}
\label{gra}
{\bf \nabla} \phi = \frac{2\pi d_e \mu_0}{\Phi_0}{\bf H}\times {\bf n} ,
\end{equation}

\noindent where ${\bf n}$ is a unit vector normal to the insulating
barrier separating the two superconducting electrodes, $\mu_0$ is
the vacuum permeability and $\Phi_0=h/2e$ is the magnetic flux
quantum. If the two superconducting films have thicknesses $d_{1,2}$
and London penetration depths $\lambda_{L1,2}$ and $t_j$ is the
barrier thickness, then the effective magnetic penetration $d_e$ is
given by\cite{wei}:

\begin{equation}
d_e=t_j + \lambda_{L1} \tanh {d_1 \over 2 \lambda_{L1}} + \lambda_{L2} \tanh {d_2 \over 2 \lambda_{L2}},
\label{d_e}
\nonumber
\end{equation}

\noindent which, in the case of thick superconducting films ($d_i
>> \lambda_{Li}$), reduces to $d_e \approx \lambda_{L1} +
\lambda_{L2}$.

\noindent Since Rowell\cite{rowell} in 1963 made the first
experimental verification of Eq.\ref{gra}, a large number of
theoretical and experimental papers have been devoted to the study
of magnetic diffraction patterns, in various Josephson junctions.
Nowadays the $I_c (H)$ curves for planar JTJs having the most common
geometrical and electrical parameters is fully understood (see, for
example, Cpt.4 and 5 of Ref.\cite{barone}). It is important to point
out that nearly all work was done with the external magnetic field
applied in the barrier plane. In fact, since Eq.\ref{gra} states
that $\phi$ is insensitive to transverse fields, this is the most
obvious choice of the magnetic field orientation.


For the reasons above, a magnetic field parallel to the barrier of
planar JTJs is applied in the practical applications of the
Josephson magnetic diffraction phenomena such as, for example, the
suppression of the d.c. Josephson effect in SIS mixers for photon
detection\cite{valery} and in specially shaped JTJs for particle
detection\cite{rogalla}, the magnetic biasing of a flux flow
oscillator\cite{nagatsuma}, the tuning of resonant fluxon
oscillators\cite{prb98}.

\section{Transverse magnetic field}

In 1975 Rosenstein and Chen\cite{rc} first reported on the effect of
a transverse magnetic field on the critical current of a JTJ with
overlap geometry. Among other things, they showed that the value of
the junction critical field $H_c$ at which the magnetic diffraction
pattern first goes to zero, changes with the inclination of the
field with respect to the barrier plane, the minimum being obtained
when the field is transverse. This was the first experimental
observation that transverse fields could be more efficient that
parallel ones in modulating the Josephson current. Soon after,
Hebard and Fulton\cite{hf} correctly interpreted the findings of
Ref.\cite{rc} in terms of stationary screening currents which
develop when a superconductor is subjected to an external magnetic
field. To better understand the mechanism through which also a
transverse field is able to modulate the critical current of a
planar JTJ, let us consider first a single isolated superconducting
film immersed in a uniform static magnetic field $H_\bot$
perpendicular to its surface. This system has received a continuous
interest over the years and here we only recall the main features.
For a deep treatment of this topic we remand to Ref.\cite{bc} and
references therein. We assume that the film thickness $d$ is larger
than its London penetration depth $\lambda_L$ and that the field
everywhere is much smaller than the critical field which would force
the film into the intermediate or normal state, i.e., that the film
is in the flux-free Meissner regime. At the top and bottom film
surfaces, the flux lines are excluded from the interior of the film
where $\bf{H}=0$. In fact, due to the screening currents
$\bf{J_s}(=\bf{\nabla} x \bf{H})$, they bend as they approach the
film surface, flow along the film surfaces, concentrate at the film
edges, and bent backward. Due to continuity, $H_n$, the component of
$\bf{H}$ normal to the surface, may be taken to be zero, while its
tangential component $H_t$ decays exponentially inside the film on
the scale of $\lambda_L$.

\noindent The knowledge of the distribution of the magnetic field
lines around the film requires a self-consistent solution of a
magnetostatic problem combining the London equation ($ \bf{H}+
\lambda_L^2 \nabla^2 \bf{H}=0$) in the superconducting film and the
fourth Maxwell equation ($ \bf{\nabla} \times \bf{H} =0$) in the
empty space around the film with boundary conditions appropriate to
the film surface geometry. This problem can be solved analytically
only for simple axially-symmetric cases such as, for example, that
of an ellipsoid of revolution with the axis of revolution parallel
to the applied field $H_\bot$\cite{Rose-Innes}.

\noindent If the film width $w$ is much less than its length, but
much greater than its thickness $d$, then we can approximate the
film as a elliptical oblate cylinder of infinite length whose
cross-section has axes $w$ and $d$; with the applied magnetic field
$H_\bot$ directed along the minor axis, then $H_t$, the component of
$\bf H$ tangent to the surface, only depends on the angle $\beta$
with respect to the minor axis: $H_t/H_\bot=1/\surd{\cos^2 \beta
+(d/w)^2\sin^2 \beta}$, whose maximum value $w/d>>1$ occurs at the
cylinder edges ($\beta=\pm \pi /2$)\cite{miller}. If the film width
$w$ is comparable to its length, the film can be approximated by a
disk whose diameter $w$ is much greater than its thickness $d$, with
the external field applied parallel to its axis. In this case, for
symmetry reasons, on the disk top and bottom surfaces the magnetic
field lines are radial and $H_t$ only depends on the distance $r$
from the disk center; it is null at the center of the disk and
increases as we move outward\cite{koba}. The surface or sheet
current density $\bf{j_s}$, defined as the screening current density
$\bf{J_s}$ integrated over the specimen thickness, equals in
magnitude $H_t$ and is everywhere orthogonal to $\bf{H}$. Numerical
simulations carried out for a $Nb$ disk having $\lambda_L=90nm$ show
that the shape of $H_t(r)$ only depends on the disk aspect ratio
$w/d$, as far as $d>>\lambda_L$. We found that at the disk border
$H_t$ is several times larger than the applied field $H_\bot$, as
shown in Fig.1 for three values of the disk aspect ratio $w/d=10$,
$100$ and $1000$, with $d=1\mu m$. Both film approximations lead to
conclude that a thin superconducting film of any geometry in a
transverse field produce a magnetic field: i) whose orientation on
the film end surfaces is parallel to surfaces themselves; ii) whose
direction near the borders of the film end surfaces is perpendicular
to the borders themselves; iii) whose strength is proportional to
the transverse applied field intensity and exceeds its value near
the borders of a film with a large aspect ratio $w/d$.

\begin{figure}[t]
\begin{center}
\epsfysize=6.0cm \epsfbox{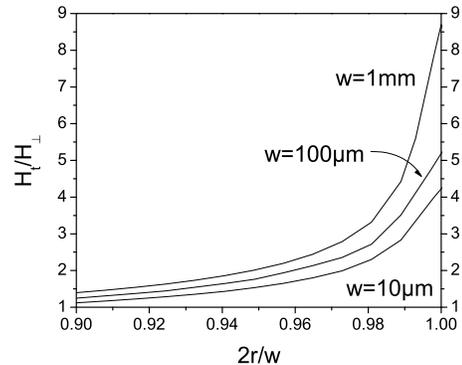}
\end{center}
\caption{Computed normalized magnetic induction field $H_t/H_\bot$
at the border of a Nb superconducting disk in a perpendicular field
$H_\bot$. The disk thickness is $d=1\mu m$ and diameters $w=10$,
$100$ and $1000 \mu m$. $2r/w$ is the normalized disk radius.}
\label{Fig1}
\end{figure}

Now, we can consider the situation in which two superconducting
films partially overlap to form a planar JTJ. If the tunnel barrier
is very transparent, then the screening currents cross the barrier
and the two films act as one single fused film. In the opposite case
the screening currents in the two films are independent on each
other and no cross talk is allowed. In the intermediate situations,
a fraction of the screening currents in one film crosses the barrier
and circulates in the other film and viceversa. It is clear that,
for a given transverse field and a given junction and electrodes
geometry, the transition from the fused to the independent films
regime can be controlled via the barrier transparency, i.e., in our
case by the Josephson current density $J_c$. The determination of
the magnetic field distribution in a system made by two
superconducting film forming a planar JTJ is a very complex task,
since it also involves the Josephson equations. The analysis becomes
even more difficult when the JTJ is biased (the distribution of the
bias current peaks near the film edges) and the junction dimensions
exceed the Josephson penetration depth $\lambda _{J}=\sqrt{\hbar
/2e\mu _{0}d_{e} J_{c}}$.

\noindent However, following Ref.\cite{miller}, we can argue that in
the independent films regime, the intensity of the in-plane magnetic
field felt by a planar JTJ placed at the borders of two
superconducting films can exceed by several times the value of the
applied transverse field. In fact, the screening currents flowing
within a depth $\lambda_{L,b}$ in the top surface of the bottom film
and those flowing within a depth $\lambda_{L,t}$ in the bottom
surface of the top film have opposite directions. Consequently the
associated magnetic fields add in the barrier plane and a more
efficient modulation of the JTJ critical current is achieved.

\section{Experiments}
In order to prove the advantages to use a
transverse field rather than an in-plane one, we have measured the
transverse magnetic diffraction patterns $I_c(H_\bot)$ of planar
high quality JTJs having different geometries, sizes and critical
current densities. The samples were placed on the axis of a superconducting coil surrounded by a Pb
shield and a cryoperm can in order to attenuate the earth magnetic field. The magnetic field produced by the coil was calculated through COMSOL MultiPhysics numerical simulations.

\subsection{Overlap type junctions}
Fig.2 compares the diffraction patterns measured in a transverse
field of two overlap-type junctions having the same geometry and
dimensions, but different critical current density $J_c$. The
junctions have been fabricated with the trilayer technique in which
the junction is realized in the window opened in an $SiO_2$
insulating layer. The thicknesses of the base, top and wiring layer
are $200$, $100$ and $400\,nm$, respectively. Details of the
fabrication process can be found in Ref.\cite{VPK}. The junction
length is $L=500\mu m$, while the width is equal to $4\mu m$. The
base and top electrode widths are $540$ and $506\mu m$,
respectively. The so called 'idle region', i.e. the overlapping of
the wiring layer onto the base electrode is about $3\,\mu m$ all
around the barrier. In Fig.2 the junction critical currents have
been normalized to their maximum values in order to make the
comparison easier. The closed circles refer to a $Nb/Al_{ox}/Nb-Nb$
tunnel junction having a critical current density $J_c=60A/cm^2$,
while the open circles refer to a $Nb/AlN/NbN-Nb$ sample with
$J_c=400A/cm^2$. As expected, considering that the field lines
associated to the screening current are perpendicular to the
electrodes edges, the shape of these $I_c$ vs. $H_\bot$ curves looks
very alike to that expected for long one-dimensional overlap-type
junctions when a uniform external field $H_\|$ is applied in the
barrier plane in the direction perpendicular to the junction length.
In fact, according to the analysis of Refs.\cite{owen,baron}, for
small field values (Meissner regime) the critical current $I_c$
decreases linearly with the applied field
$I_c(H_\|)=I_c(0)(1-|H_\||/H_c^{\|})$, where $H_c^\|$ is the
critical field at which $I_c$ would vanish if flux quanta did not
start to enter the junction barrier. The skewness seen in the
experimental $I_c - H_\bot$ curves (being larger for the sample
having the larger $J_c$) is due to the self-field produced by the
bias current $I$ flowing in a close-by superconducting strip in the
chip circuitry. The skewness does not prevent us from measuring the
junction critical critical fields $H_c^\bot$ determined by linearly
extrapolating the branches starting at maximum critical current to
$I=0$ (dotted lines in Fig.1); in fact, when $I=0$, the bias current
self-field effects vanish. The critical field $H_c^\|$ for a long
one-dimensional junction in presence of a in-plane external field
applied perpendicular to the long junction dimension $L$ can be
expressed as the sum of two terms:

\begin{equation}
{H_c^{\|}}=H_{c}^{F}+H_{c}^{FP} =\frac {\Phi_0}{\mu_0 d_e L} +
2\lambda_J J_c
\nonumber
\end{equation}

\begin{equation}
=  \frac {\Phi_0}{ \mu_0 d_e } (\frac {1}{L}+\frac
{1}{\pi \lambda_J})= \frac {\Phi_0}{\mu_0 d_e L} + \sqrt{\frac{2
\Phi_0 J_c}{\pi \mu_0 d_e}}.\label{Hco}
\end{equation}

The first term $H_{c}^{F}$ in Eq.\ref{Hco} dominates in low $J_c$
samples for which $L<<\pi \lambda_j$ and corresponds to the critical
field of a point-like junction which exhibits the well known
Fraunhofer diffraction pattern. The second term  $H_{c}^{FP} \propto
\sqrt{{J_c}/{d_e}}$ becomes dominant in the high $J_c$ regime when
$L>>\pi \lambda_j$. It was first introduced by Ferrell and
Prange\cite{ferrell} in order to describe the self-field limiting
effects in long inline-type junctions.

\begin{figure}[t]
\begin{center}
\epsfysize=6.0cm \epsfbox{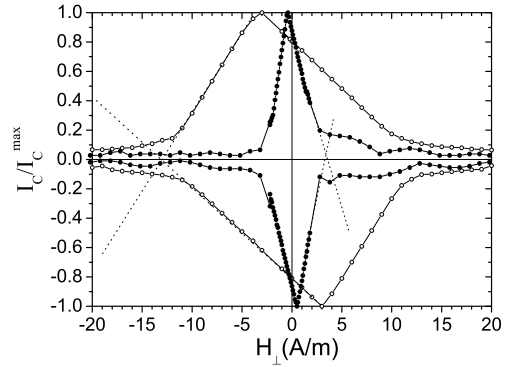}
\end{center}
\caption{Magnetic diffraction patterns measured in a transverse
field $H_{\bot}$ of two overlap-type junctions $L \times W = 500
\times 4 \mu m^2$. The closed circles refer to a $Nb/Al_{ox}/Nb$ JTJ
having $J_c=60A/cm^2$, $d_e=180nm$ and $\lambda_J \sim 50 \mu m$.
The open circles refer to a $Nb/AlN/NbN$ with $J_c=400A/cm^2$,
$d_e=240nm$ and $\lambda_J \sim 20 \mu m$. The dotted lines show the
procedure used to determine the junction critical field
$H_{c}^{\bot}$.} \label{Fig2}
\end{figure}

\begin{table}[b]
\begin{tabular*}{0.45\textwidth}%
{@{\extracolsep{\fill}}cccccc}
Sample & $J_c$ & $d_e$ & $L/\lambda_J$ & $H_{c}^{\bot}$& $H_{c}^{\|}=H_{c}^{F}+H_{c}^{FP}$\\
 & $A/cm^2$ & $nm$ &  & $A/m$& $A/m$\\
\hline  
$Nb/Al_{ox}/Nb$ & $60$  & $180$  & $\sim 10$ & $3.2$ & $36+58=94$\\
$Nb/AlN/NbN$  & $410$  & $240$  & $\sim 25$ & $13$ & $36+131=167$\\
\end{tabular*}
\caption{Relevant parameters for two overlap-type Josephson tunnel
junctions whose transverse magnetic diffraction patterns are shown
in Fig.2.}
\end{table}

Table.1 reports the values of $J_c$ and $d_e$ used to predict the
critical field $H_{c}^{\|}$ from to Eq.\ref{Hco}. The last two
columns allow the comparison between $H_{c}^{\bot}$ and
$H_{c}^{\|}$. We observe that for both samples
$H_{c}^{\|}>H_{c}^{\bot}$, and the ratio $H_{c}^{\|}/H_{c}^{\bot}$
changes from about $30$ to about $13$ when we move from the low to
the high $J_c$ junction. The data in Table.1 unambiguously indicate
that a transverse field can be more effective than a in-plane one
and, remembering that the two samples have the same geometrical
details, the effect of a transverse field weakens as the junction
transparency increases.

\subsection{Annular junctions}
Further, we have measured the static
properties of several $Nb/Al_{ox}/Nb-Nb$ annular JTJs having the
same critical current density $J_c=60A/cm^2$ ($\lambda_J=50\mu m$),
the same width $\Delta R=4\mu m$, but different radius $R$ ranging
from $80$ to $500 \mu m$. The fabrication details are the same as
those for the overlap-type JTJs discussed previously. The
diffraction patterns in transverse magnetic field will be reported
elsewhere\cite{next}; here we focus our interest on the values of
the critical fields.

\noindent The analogous of Eq.\ref{Hco} for a one-dimensional
annular junction having radius $R$ in an in-plane field
is\cite{br96}:
\begin{equation}
 {H_c^{\|}}= H_{c}^{B}+H_{c}^{MM}= 2.404 \frac {\Phi_0}{2 \pi \mu_0 d_e R} + R J_c.
\label{Hca}
\end{equation}

\noindent Again we have a contribution $H_{c}^{B}$, independent on
the Josephson current density $J_c$, typical of small and
intermediate radius annular JTJs immersed in a uniform in-plane
magnetic field which results in a periodic radial field $H_r(\theta)
\propto \cos \theta$ felt by the junction\cite{lom,prb96}. In such
case, the $I_c$ vs. $H_{\|}$ curve follows a Bessel, rather than a
Fraunhofer, behavior (2.404 is the argument corresponding to the
first minimum of the zero-order Bessel function). The second term
$H_{c}^{MM}=R J_c$ in Eq.\ref{Hca} was numerically found by
Martucciello and Monaco\cite{br96}; considering that
$H_{c}^{MM}/H_{c}^{B}=(R/\lambda_J)^2/2.404$, it becomes dominant
when $R>>R_m=\sqrt{2.404} \lambda_J$\cite{prb96,prb98}. For given
$d_e$ and $J_c$, Eq.\ref{Hca} has a minimum when $R=R_m$ and
linearly increase with $R$ when $R>>R_m$. The last two columns of
the Table.2 report, respectively, the transverse critical field
$H_{c}^{\bot}$ measured for four annular junctions having different
radii and the expected parallel critical field $H_{c}^{\|}$
according to Eq.\ref{Hca} with $J_c=60A/cm^2$ and $d_e=180nm$.

\begin{table}[t]
\begin{center}  
\begin{tabular*}{0.45\textwidth}%
{@{\extracolsep{\fill}}cccc}
$R$&$R/\lambda_J$ &$H_{c}^{\bot}$& $H_{c}^{\|}=H_{c}^{B}+H_{c}^{MM}$\\
$mm$ & &$A/m$& $A/m$\\
\hline  
$0.08$  &$\sim1.6$&  $2.7$ & $18+48=66$\\
$0.25$  &$\sim5$& $1.5$ & $6+150=156$\\
$0.32$  &$\sim6.4$& $1.3$& $4+192=196$\\
$0.50$  &$\sim10$& $1.2$& $3+300=303$\\
\end{tabular*}
\end{center}
\caption{Relevant parameters of the $Nb/Al_{ox}/Nb$ annular Josephson tunnel junctions used in the experiments.}
\end{table}

For all samples we observe once again that
$H_{c}^{\|}>H_{c}^{\bot}$, and that the ratio
$H_{c}^{\|}/H_{c}^{\bot}$ changes from about $25$ to about $250$ when
we increase the ring diameter, i.e. the top and bottom film widths,
confirming that the effect of a transverse field strengthens as the
electrode widths increases. In particular, for the three largest
rings, having the so called Lyngby geometry\cite{davidson}, i.e. the
base and top electrode widths match the ring diameter, this ratio is
proportional to $R$. [The smallest ring ($R/\lambda_J \approx 1.6$)
has the base electrode width equal to $540 \mu m$ that is
considerably larger than the ring diameter $160 \mu m$.]

\section{Concluding remarks}

\noindent We have measured the transverse magnetic diffraction
pattern of planar JTJs having different geometries, sizes and
barrier transparencies. Our measurements clearly indicate that a
magnetic field is more effective to modulate the junction critical
current $I_c$ when applied perpendicularly (rather than parallel) to
the junction plane provided the JTJ is fabricated close to the
borders of superconducting films with a large aspect ratio. This is
due to screening (or Meissner) currents induced by the transverse
field that circulate mainly on the film surface borders which in
turn behave as intrinsic control lines. We suggest that a transverse
magnetic field can be usefully exploited in those applications where
the Josephson critical current and the Fiske resonances need to be
suppressed.


\end{document}